\documentclass[english,notitlepage,prl,twocolumn]{revtex4-2}
\usepackage[T1]{fontenc}
\usepackage[latin9]{inputenc}
\usepackage{amsmath}
\usepackage{amssymb}
\usepackage{graphicx}

\makeatletter
\usepackage[colorlinks,urlcolor=cyan,citecolor=blue,linkcolor=magenta]{hyperref}
\usepackage{xcolor}
\newcommand{\prlsection}[1]{%
  \noindent\textit{\textcolor{blue}{#1}}---%
}

\makeatother

\usepackage{babel}
\begin{document}
\title{Breathing Modes as a Probe of Energy Fluctuations in a Unitary Fermi
Gas}
\author{Shi-Guo Peng$^{1,2,4}$, Jin Min$^{3}$, Kaijun Jiang$^{2,5}$}
\email{kjjiang@wipm.ac.cn}

\affiliation{$^{1}$Center for Theoretical Physics, Hainan University, Haikou 570228,
China}
\affiliation{$^{2}$Innovation Academy for Precision Measurement Science and Technology,
Chinese Academy of Sciences, Wuhan 430071, China}
\affiliation{$^{3}$State Key Laboratory for Mesoscopic Physics, School of Physics,
Frontiers Science Center for Nano-optoelectronics, Collaborative Innovation
Center of Quantum Matter, Peking University, Beijing 100871, China}
\affiliation{$^{4}$School of Physics and Optoelectronic Engineering, Hainan University,
Haikou 570228, China}
\affiliation{$^{5}$Wuhan Institute of Quantum Technology, Wuhan 430206, China}
\date{\today}
\begin{abstract}
Directly accessing energy fluctuations in interacting quantum many-body
systems remains a long-standing challenge, especially far from equilibrium.
Here we show that in scale-invariant quantum gases with SO$(2,1)$
dynamical symmetry, the amplitude of the breathing mode provides a
direct and quantitative probe of energy fluctuations. We establish
an exact and universal relation between the oscillation amplitude
and the energy fluctuation, with a dimensionless ratio fixed solely
by the Bargmann index $k$, which labels the irreducible representation
of the underlying SU$(1,1)$ algebra and thereby determines the structure
of the many-body spectrum and dynamics. As a consequence, this relation
is fully dictated by symmetry and remains independent of microscopic
details and excitation protocols. Furthermore, we show that the excitation
of breathing-mode states follows a universal statistical distribution
governed by a single parameter, independent of the specific driving
protocol. Our findings demonstrate that energy fluctuations, typically
encoded in the many-body spectrum, can be directly accessed through
collective dynamics, offering a symmetry-based route to probe nonequilibrium
energy statistics in strongly interacting quantum systems.
\end{abstract}
\maketitle
\prlsection{Introduction}Directly accessing energy fluctuations in
an interacting quantum many-body system remains a long-standing challenge,
yet such fluctuations play a central role in statistical physics and
quantum dynamics. In equilibrium, energy fluctuations encode fundamental
thermodynamic properties through the fluctuation--dissipation theorem
and determine quantities such as heat capacity and critical response
near phase transitions \citep{Landau1980S,Pathria1972S,Callen1951I,Kubo1966T}.
Far from equilibrium, however, energy fluctuations become even more
central: they characterize work and heat statistics, entropy production,
and universal fluctuation relations such as the Jarzynski equality
and Crooks fluctuation theorem \citep{Jarzynski1997N,Crooks1999E,Mukamel2003Q,Talkner2007F,Esposito2009N,Campisi2011Q,Hanggi2015T}.
More broadly, energy fluctuations govern relaxation, thermalization,
and dynamical responses in nonequilibrium quantum many-body systems,
providing key insights into irreversibility and quantum thermodynamics
\citep{Esposito2009N,Vinjanampathy2016Q}. Accessing energy fluctuations
is therefore essential for understanding the physics of nonequilibrium
quantum matter.

Yet, while the mean energy of a quantum system is routinely accessible,
determining higher moments of the energy distribution remains far
more challenging. In quantum systems, energy fluctuations are encoded
in the many-body spectrum and cannot generally be extracted from a
single observable measurement. Existing approaches therefore rely
on protocols based on full counting statistics \citep{Talkner2007F,Esposito2009N,Campisi2011Q},
which reconstruct the characteristic function of the work or energy
distribution, often implemented through interferometric schemes involving
an ancillary qubit \citep{Dorner2013E,Mazzola2013M}. Alternatively,
energy fluctuations can be inferred from reconstruction methods based
on transition probabilities between many-body eigenstates \citep{Herrera2021E}.
While these approaches establish conceptually powerful paradigms,
their implementation becomes increasingly demanding for large interacting
many-body systems. Identifying simple and scalable probes that convert
microscopic energy fluctuations into directly measurable observables
therefore remains an open problem.

Here we demonstrate that in scale-invariant quantum gases with SO$(2,1)$
symmetry \citep{Pitaevskii1997B,Werner2006U,Wang2024S,Sun2025P,Yan2025E},
the amplitude of the breathing mode $\mathcal{A}$ provides a direct
and quantitative measure of energy fluctuation $\Delta E$. We establish
an exact relation
\begin{equation}
\frac{\Delta E/\hbar\omega}{\mathcal{A}/a^{2}_{\text{ho}}}=\frac{1}{\sqrt{2k}},\label{eq:UniversalRatio}
\end{equation}
where $a_{\text{ho}}=\sqrt{\hbar/m\omega}$ is the harmonic length
for a trap with frequency $\omega$ and atomic mass $m$. The parameter
$k$ is the Bargmann index that labels the irreducible representation
of the SU$(1,1)$ algebra associated with the SO$(2,1)$ dynamical
symmetry. It thus encodes the symmetry-determined structure of the
many-body state and renders the above dimensionless relation universal,
independent of microscopic details, excitation protocols and system
parameters. Crucially, while collective modes are typically sensitive
only to average thermodynamic quantities, we show that the breathing
amplitude is uniquely sensitive to energy fluctuations, which are
otherwise inaccessible without full spectral information. This establishes
a symmetry-protected mapping between a microscopic quantity (energy
fluctuation) and a macroscopic observable (collective oscillation
amplitude). More generally, our result establishes a symmetry-based
paradigm in which complex spectral properties are directly reflected
in collective dynamics, enabling experimental access to quantum fluctuations
without full spectral reconstruction.

\emph{\prlsection{Quasiparticle structure}}The breathing dynamics
of a scale-invariant quantum gas in a harmonic trap is governed by
the SO$(2,1)$ symmetry generated by the free-space Hamiltonian $\hat{H}_{0}=\sum_{i}{\bf p}^{2}_{i}/2m+\hat{V}_{\text{int}}$,
the dilatation operator $\hat{D}=\sum_{i}\left({\bf r}_{i}\cdot{\bf p}_{i}+{\bf p}_{i}\cdot{\bf r}_{i}\right)/2$,
and the special conformal operator $\hat{K}=\sum_{i}mr^{2}_{i}/2$
\citep{Pitaevskii1997B,Werner2006U}. Here ${\bf r}_{i}$ and ${\bf p}_{i}$
denote the coordinate and momentum operators of the $i$th atom, and
$\hat{V}_{\text{int}}$ describes the interatomic interactions. The
SO$(2,1)$ symmetry requires scale-invariant interactions, realized
in a Fermi gas at the unitary limit where the scattering length diverges
and no intrinsic length scale remains \citep{Peng2023D}. In a harmonic
trap with frequency $\omega$, these generators can be reorganized
into the SU$(1,1)$ algebra with generators $\hat{L}_{0}=(\hat{H}_{0}+\omega^{2}\hat{K})/2\omega$
and $\hat{L}_{\pm}=(\hat{H}_{0}-\omega^{2}\hat{K}\pm i\omega\hat{D})/2\sqrt{2}\omega$,
which satisfy the commutation relations $[\hat{L}_{0},\hat{L}_{\pm}]=\pm\hbar\hat{L}_{\pm}$
and $[\hat{L}_{+},\hat{L}_{-}]=-\hbar\hat{L}_{0}$. The irreducible
representations of this algebra are labeled by the Casimir operator
$\hat{C}=\hat{\mathcal{H}}^{2}-\left(2\omega\right)^{2}(\hat{L}_{+}\hat{L}_{-}+\hat{L}_{-}\hat{L}_{+})$,
which is symmetric under the algebra and thus conserved throughout
the dynamics. The Hilbert space can thus be decomposed into sectors
with fixed Casimir eigenvalue, and the dynamics under a time-dependent
trap remains confined within a single such sector. Each sector corresponds
to an irreducible representation labeled by the Bargmann index $k$,
which is directly related to the Casimir eigenvalue and determines
the lowest energy of the breathing tower $\epsilon_{g}=(2\hbar\omega)k$
\citep{CasimirEigenvalue}. Within this representation the many-body
spectrum forms a ladder of states $\left|k,n\right\rangle \propto(\hat{L}_{+})^{n}\left|k,0\right\rangle $
with energies $\epsilon_{n}=\epsilon_{g}+2n\hbar\omega$ \citep{Maki2019Q,Maki2020F,Maki2022D}.
The equally spaced structure of this breathing spectrum naturally
suggests a quasiparticle description \citep{Werner2006U}. Introducing
bosonic operators $\{\hat{b}^{\dagger},\hat{b}\}$ satisfying $[\hat{b},\hat{b}^{\dagger}]=1$,
we define their action on the tower states through the quasiparticle
number operator $\hat{n}=\hat{b}^{\dagger}\hat{b}$. In this representation
the SU$\left(1,1\right)$ generators take the form 
\begin{equation}
\hat{L}_{0}=\hbar\left(\hat{n}+k\right),\;\hat{L}_{+}=\hat{L}^{\dagger}_{-}=\frac{\hbar}{\sqrt{2}}\hat{b}^{\dagger}\sqrt{\hat{n}+2k}.\label{eq:QPOperators}
\end{equation}
The total Hamiltonian of a trapped unitary Fermi gas can then be written
simply as $\hat{\mathcal{H}}=2\omega\hat{L}_{0}=2\hbar\omega(\hat{n}+k)$.
This representation provides a transparent physical picture. The lowest-energy
state $\left|k,0\right\rangle $ acts as a quasiparticle vacuum, while
the tower states correspond to excitations of a collective breathing
quasiparticle with universal energy spacing $2\hbar\omega$. The breathing
dynamics can therefore be interpreted as the creation and annihilation
of quasiparticles generated by $\hat{b}^{\dagger}$ and $\hat{b}$.
In this language, the universality of the breathing frequency $2\omega$
is not accidental but follows directly from the ladder structure of
the underlying SO$\left(2,1\right)$ representation. This quasiparticle
picture thus provides a natural framework for describing the nonequilibrium
breathing dynamics discussed below.

\prlsection{Breathing-mode excitation}The nonequilibrium excitation
of the breathing mode admits an exact description due to the underlying
SO$(2,1)$ symmetry. We consider a unitary Fermi gas trapped in a
general time-dependent harmonic trap with frequency $\omega(t)$,
where the time-evolution operator satisfies $i\hbar\partial_{t}\hat{U}=\hat{\mathcal{H}}\hat{U}$
with $\hat{\mathcal{H}}(\omega)=\hat{H}_{0}+\omega^{2}(t)\hat{K}$.
Since $\hat{\mathcal{H}}(\omega)$ remains a linear combination of
the SO$(2,1)$ generators, the dynamics is exactly confined within
the SU$(1,1)$ algebra and the time-evolution operator can be written
in disentangled decomposition as \citep{Zhang2022Q,Ban1993D}
\begin{equation}
\hat{U}\left(t\right)=e^{\sqrt{2}\zeta_{+}\left(t\right)\hat{L}_{+}/\hbar}e^{\hat{L}_{0}\ln\left[1-\left|\zeta_{+}\left(t\right)\right|^{2}\right]/\hbar}e^{-\sqrt{2}\zeta^{*}_{+}\left(t\right)\hat{L}_{-}/\hbar}.
\end{equation}
The complex parameter $\zeta_{+}$ obeys the Riccati equation
\begin{equation}
i\frac{d}{dt}\zeta_{+}=\frac{1}{2}\alpha_{-}+\alpha_{+}\zeta_{+}+\frac{1}{2}\alpha_{-}\zeta^{2}_{+}\label{eq:RiccatiEq}
\end{equation}
 with $\alpha_{\pm}\left(t\right)=\omega_{0}[1\pm\omega^{2}\left(t\right)/\omega^{2}_{0}]$
and initial trap frequency $\omega_{0}$, which can be parameterized
as $\zeta_{+}=e^{i\theta}\tanh s$. This reveals that the full many-body
evolution governed by $\hat{U}\left(t\right)$ is equivalent to a
generalized SU$\left(1,1\right)$ displacement (squeezing) transformation
$\hat{U}_{D}(\xi)=e^{\sqrt{2}(\xi\hat{L}_{+}-\xi^{*}\hat{L}_{-})/\hbar}$
with $\xi=se^{i\theta}$. Remarkably, the entire many-body dynamics
is exactly reduced to the evolution of a single complex parameter
$\xi(t)$. The breathing excitation thus corresponds to a symmetry-constraint
squeezing of the many-body wave function rather than a generic redistribution
over exponentially many states. In contrast to noninteracting systems,
where the generator of the transformation is quadratic in bosonic
operators, the square-root structure of $\hat{L}_{\pm}$ reflects
the intrinsic many-body correlations encoded in the SU$\left(1,1\right)$
representation, implying that the elementary excitations are collective
in nature. 

The SU$(1,1)$ displacement acts nontrivially on the breathing quasiparticles
and induces a nonlinear transformation of the quasiparticle operators
$\hat{b}(t)=\hat{U}^{\dagger}_{D}(\xi)\hat{b}_{0}\hat{U}_{D}(\xi)$,
where $\hat{b}_{0}$ annihilates the quasiparticles associated with
the initial frequency $\omega_{0}$. Using the transformation of Eq.(\ref{eq:QPOperators}),
one obtains \citep{SM}
\begin{multline}
\hat{n}\left(t\right)+k=\cosh\left(2s\right)\left(\hat{n}_{0}+k\right)\\
+\frac{e^{i\theta}}{2}\sinh\left(2s\right)\hat{B}^{\dagger}_{0}+\frac{e^{-i\theta}}{2}\sinh\left(2s\right)\hat{B}_{0},\label{eq:TimeEvolutionOfN}
\end{multline}
\begin{multline}
\hat{B}\left(t\right)=e^{i\theta}\sinh\left(2s\right)\left(\hat{n}_{0}+k\right)\\
+e^{i2\theta}\sinh^{2}\left(s\right)\hat{B}^{\dagger}_{0}+\cosh^{2}\left(s\right)\hat{B}_{0},\label{eq:TimeEvolutionOfB}
\end{multline}
where $\hat{B}_{0}=\sqrt{\hat{n}_{0}+2k}\hat{b}_{0}$ with $\hat{n}_{0}=\hat{b}^{\dagger}_{0}\hat{b}_{0}$,
and $\hat{B}(t)=\hat{U}^{\dagger}_{D}(\xi)\hat{B}_{0}\hat{U}_{D}(\xi)$.
In contrast to the linear Bogoliubov transformation familiar from
non-interacting systems and quantum optics \citep{Xin2021R}, this
evolution is intrinsically nonlinear, reflecting the interacting many-body
nature encoded in the SU$(1,1)$ representation. For an initial quasiparticle
vacuum $\hat{b}_{0}\left|0;\omega_{0}\right\rangle =0$, the displacement
generates quasiparticles within the same irreducible representation
(fixed by $k$). The quasiparticle number $\mathcal{N}\left(t\right)=\left\langle \hat{n}\left(t\right)\right\rangle _{0}$
is obtained by taking the expectation value with respect to the initial
quasiparticle vacuum $\left|0;\omega_{0}\right\rangle $. Using the
transformation (\ref{eq:TimeEvolutionOfN}), we obtain the exact result
$\mathcal{N}(t)=k[\cosh(2s)-1]$. The quasiparticle number also exhibits
enhanced fluctuations
\begin{equation}
\Delta\mathcal{N}\left(t\right)=\sqrt{\left\langle \hat{n}^{2}\left(t\right)\right\rangle _{0}-\left\langle \hat{n}\left(t\right)\right\rangle ^{2}_{0}}=\sqrt{\frac{k}{2}}\sinh\left(2s\right).
\end{equation}
This result deviates strikingly from independent-particle statistics.
For uncorrelated excitation processes one expects Poissonian scaling
$\Delta\mathcal{N}\sim\sqrt{\mathcal{N}}$, whereas here one finds
$\Delta\mathcal{N}\sim\mathcal{N}$, corresponding to super-Poissonian
fluctuations \citep{Scully1997Q}. This enhanced fluctuation does
not arise from randomness but reflects the collective nature of the
excitation process: quasiparticles are generated through a coherent
squeezing dynamics rather than as independent events. Importantly,
both the mean quasiparticle number and its fluctuation are governed
by the same single parameter $s\left(t\right)$, showing that the
entire nonequilibrium excitation dynamics is controlled by a single
SU$(1,1)$ squeezing coordinate. 

\emph{\prlsection{The amplitude-energy-fluctuation relation}}After
the excitation stage with duration $t_{1}$, the system undergoes
a free breathing oscillation in a static trap with fixed frequency
$\omega$. The subsequent dynamics is governed by $\hat{\mathcal{H}}(\omega)=\hat{H}_{0}+\omega^{2}\hat{K}$,
which evolves in the same SU$(1,1)$ representation and takes the
diagonal form $\hat{\mathcal{H}}(\omega)=2\hbar\omega(\hat{n}+k)$,
where $\hat{n}$ is the breathing-mode quasiparticle number operator
defined with respect to the final trap frequency $\omega$. Since
$[\hat{n},\hat{\mathcal{H}}(\omega)]=0$, the quasiparticle number
is conserved during the free evolution. Consequently, both its mean
value $\mathcal{N}=\langle\hat{n}\rangle$ and fluctuation are fully
determined by the excitation stage. 

The excitation generates a displaced state described by the squeezing
operator $\hat{U}_{D}(\xi)$. After the excitation is switched off,
the system is naturally described in the quasiparticle basis associated
with the final trap frequency $\omega$, which differs from that of
the initial Hamiltonian. The relation between the two quasiparticle
descriptions is established by the scale transformation $\hat{U}_{S}(v)=e^{iv\hat{D}/\hbar}$
with $e^{2v}=\omega/\omega_{0}$, which maps the Hamiltonians as $\hat{\mathcal{H}}(\omega)=\hat{U}_{S}(v)\hat{\mathcal{H}}(\omega_{0})\hat{U}^{\dagger}_{S}(v)$.
Importantly, this transformation does not create excitations, but
mixes quasiparticle creation and annihilation operators, thereby reshuffling
the excitation content when expressed in the new basis. Because both
$\hat{U}_{D}(\xi)$ and $\hat{U}_{S}(v)$ belong to the same SU$(1,1)$
group, their combined action remains within the same manifold of squeezed
states. As a result, the entire excitation protocol is governed by
a single effective squeezing amplitude $\mathcal{S}_{\text{eff}}$
defined by \citep{SM}
\begin{equation}
\cosh\left(2\mathcal{S}_{\text{eff}}\right)=\cosh\left(2s\right)\cosh\left(2v\right)-\sinh\left(2s\right)\sinh\left(2v\right)\cos\theta.
\end{equation}
This reduction implies that both the mean quasiparticle number and
its fluctuation are controlled by a single parameter $\mathcal{S}_{\text{eff}}$,
\begin{equation}
\mathcal{N}=k\left[\cosh\left(2\mathcal{S}_{\text{eff}}\right)-1\right],\;\Delta\mathcal{N}=\sqrt{\frac{k}{2}}\sinh\left(2\mathcal{S}_{\text{eff}}\right),
\end{equation}
showing that the excitation strength and its fluctuation are not independent,
but constrained by the underlying SO$(2,1)$ symmetry.

We now connect these quantities to the experimentally measurable cloud
size. The breathing dynamics is governed by the operator $\hat{K}$,
which does not commute with the Hamiltonian and therefore exhibits
oscillatory evolution. A straightforward evaluation yields
\begin{equation}
\left\langle r^{2}\right\rangle _{\tau}=\frac{E}{m\omega^{2}}-\mathcal{A}\cos\left(2\omega\tau-\delta\right),
\end{equation}
where $\tau=t-t_{1}$, $\delta$ is a phase determined by the squeezing
parameters $(s,\theta,v)$ \citep{SM}. The equilibrium position is
fixed by the total energy $E=2\hbar\omega(\mathcal{N}+k)$, recovering
the well-known result that the mean cloud size directly measures the
total energy in scale-invariant systems \citep{Wang2024S,Sun2025P,Yan2025E}.
The oscillation amplitude is given by $\mathcal{A}/a^{2}_{\text{ho}}=2\sqrt{2k}\Delta\mathcal{N}$,
showing that while the mean radius probes the average energy, the
oscillation amplitude is governed entirely by the quasiparticle number
fluctuation. This leads to a direct relation between the amplitude
and energy fluctuation $\Delta E=2\hbar\omega\Delta\mathcal{N}$ given
by Eq.(\ref{eq:UniversalRatio}). Remarkably, in dimensionless form,
the ratio between energy fluctuation and oscillation amplitude is
universally fixed by the Bargmann index $k$, which labels the irreducible
representation of the underlying symmetry, and is therefore independent
of microscopic details, excitation protocols, and trap parameters
within a given sector. This identifies a symmetry-protected and experimentally
accessible mapping between a collective observable and quantum energy
fluctuations, arising solely from the underlying SO$(2,1)$ dynamical
symmetry. 

\begin{figure}
\includegraphics[width=1\columnwidth]{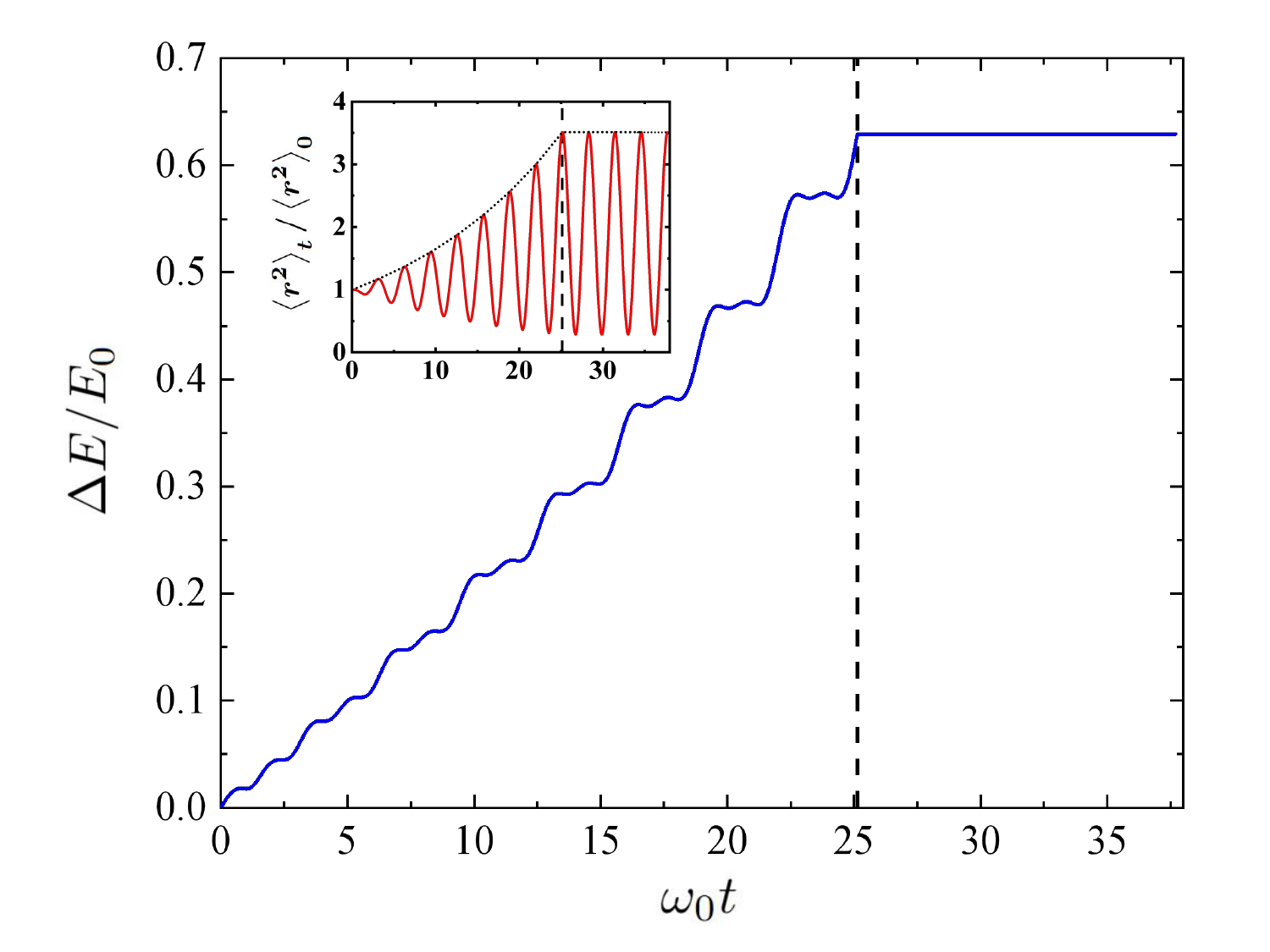}

\caption{(Color online) Time evolution of the energy fluctuation under resonant
modulation. The vertical dashed line denotes the end of the modulation
stage, separating the driven and free-evolution regimes. During modulation,
the energy fluctuation increases due to continuous energy injection,
while it remains constant after the driving is switched off. The inset
shows the corresponding evolution of the cloud size $\langle r^{2}\rangle_{t}/\langle r^{2}\rangle_{0}$,
where the oscillation amplitude grows during the excitation stage
and stays constant thereafter. Here, $E_{0}$ and $\langle r^{2}\rangle_{0}$
are the initial energy and cloud size. The trap frequency is modulated
as $\omega^{2}(t)=\omega^{2}_{0}[1+\beta\sin(2\omega_{0}t)]$, resonant
with the breathing mode at frequency $2\omega_{0}$, with modulation
strength $\beta=0.1$.}

\label{fig:EnergyFluctuation}
\end{figure}

\emph{\prlsection{Universality across excitation protocols}}To demonstrate
the universality of the amplitude-energy-fluctuation relation, we
consider two experimentally relevant excitation protocols: resonant
modulation and sudden quench. We first discuss the resonant modulation,
where the trap is driven as $\omega^{2}(t)=\omega^{2}_{0}[1+\beta\sin(2\omega_{0}t)]$,
resonantly coupling to the breathing mode. Within the SU$(1,1)$ framework,
the entire dynamics reduces to a parametric amplification process
controlled by a single complex parameter $\zeta_{+}=e^{i\theta}\tanh s$,
which evolves according to the Riccati equation (\ref{eq:RiccatiEq}).
As a result, breathing quasiparticles are generated coherently, leading
to a rapid growth of both the quasiparticle number and its fluctuation
$\Delta\mathcal{N}\sim\mathcal{N}$. The evolution of the energy fluctuation
during the modulation is shown in Fig.\ref{fig:EnergyFluctuation}.
While the overall trend exhibits a clear increase due to resonant
energy injection, the dynamics within each modulation cycle reveals
a nontrivial oscillatory structure. In particular, the growth of the
energy fluctuation is intermittently slowed down, forming plateau-like
features that become increasingly pronounced at later times. This
behavior originates from the interplay between the external driving
and the intrinsic breathing dynamics. The instantaneous rate of energy
change is determined by 
\begin{equation}
\frac{dE}{dt}=\left\langle \frac{\partial\hat{\mathcal{H}}}{\partial t}\right\rangle \propto\dot{\omega}\left(t\right)\left\langle r^{2}\right\rangle _{t},
\end{equation}
while $\langle r^{2}\rangle_{t}$ itself oscillates at the breathing
frequency $2\omega_{0}$. As a consequence, the work done by the trap
is not monotonic within each cycle: depending on the relative phase
between the modulation and the breathing motion, the system alternates
between energy absorption and partial energy release. This leads to
the oscillatory plateau structure observed in the the evolution of
the energy fluctuation. 

As excitation proceeds, the amplitude of the breathing motion increases,
driving the system into a strongly nonlinear regime. In this regime,
the phase relation between the drive and the collective motion becomes
increasingly sensitive, reducing the efficiency of energy absorption
during parts of the cycle. Physically, this reflects the fact that
when the cloud is strongly compressed, further compression becomes
energetically costly, effectively limiting the work that can be done
by the external drive. As a result, the plateau structures become
more pronounced at larger times. The small oscillations within each
plateau can be understood as a manifestation of energy backflow: during
certain phases of the cycle, the system transiently returns energy
to the driving field. This effect is a characteristic feature of parametric
resonance in interacting systems and is naturally captured by the
SU$(1,1)$ dynamics, where the squeezing evolution combines exponential
growth with intrinsic phase oscillations. After the modulation is
switched off, the system undergoes free evolution in the static trap,
where the quasiparticle number is conserved and the energy fluctuation
remains constant. The corresponding breathing dynamics of the cloud
size is shown in the inset of Fig.\ref{fig:EnergyFluctuation}.

For the quench process, where the trap frequency is suddenly changed
from $\omega_{0}$ to a final value $\omega$, the system is driven
out of equilibrium and subsequently undergoes free breathing oscillations.
In this case, the excitation is entirely determined by the mismatch
between the initial state and the post-quench Hamiltonian. As a result,
the quasiparticle number and its fluctuation, once generated, remain
conserved during the subsequent evolution, reflecting the conservation
of energy and its fluctuation. Within the SU$(1,1)$ framework, this
process admits a particular simple description. Since the quench involves
no dynamical squeezing ($s=0$), the excitation reduces to a scale
transformation generated by $\hat{U}_{S}(v)$, which connects the
initial and final Hamiltonians via $\hat{\mathcal{H}}(\omega)=\hat{U}_{S}(v)\hat{\mathcal{H}}(\omega_{0})\hat{U}^{\dagger}_{S}(v)$,
with $e^{2v}=\omega/\omega_{0}$. The post-quench state thus remains
within the same SU$(1,1)$ manifold and is fully described by a single
effective parameter $\mathcal{S}_{\text{eff}}=v$. The energy fluctuation
and the breathing amplitude then follow directly from this parameter.

\begin{figure}
\includegraphics[width=1\columnwidth]{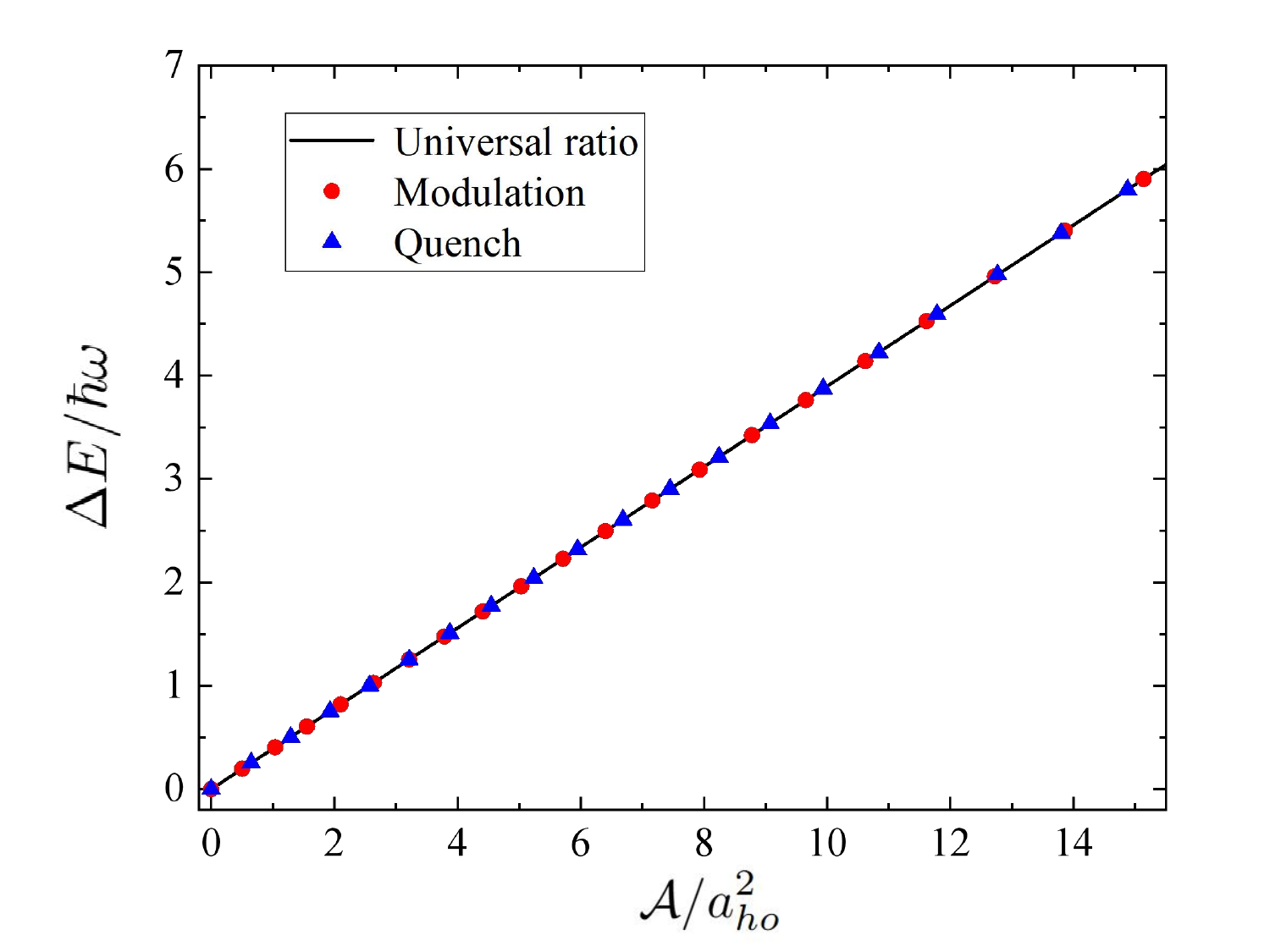}

\caption{(Color online) Energy fluctuation $\Delta E/\hbar\omega$ versus breathing
amplitude $\mathcal{A}/a^{2}_{\text{ho}}$ for quench and resonant
modulation protocols. All numerical data collapse onto a single line
with slope $1/\sqrt{2k}$, demonstrating the universal amplitude-energy-fluctuation
relation (\ref{eq:UniversalRatio}), which is independent of the specific
excitation protocol and microscopic details. Here, $k$ is the Bargmann
index that labels the irreducible representation of the SU$(1,1)$
algebra associated with the SO$(2,1)$ symmetry, and $a_{\text{ho}}=\sqrt{\hbar/m\omega}$
is the harmonic length.}

\label{fig:UniversalRatio}
\end{figure}

In Fig.\ref{fig:UniversalRatio}, we plot the energy fluctuation $\Delta E/\hbar\omega$
as a function of the breathing amplitude $\mathcal{A}/a^{2}_{\text{ho}}$
for the quench protocol, and compare it with the results from resonant
modulation. While the two protocols generate excitations through entirely
different mechanisms---the quench through a sudden projection and
the modulation through continuous parametric driving---their outcomes
exhibit a remarkable universality. For the quench, the excitation
strength is controlled by the ratio $\omega/\omega_{0}$, whereas
for the resonant modulation it is governed by the number of driving
cycles. Despite these differences, all data collapse onto a single
straight line with slope $1/\sqrt{2k}$, with $k$ determined by underlying
symmetry and the structure of its irreducible representations. This
shows that the experimentally measurable slope directly probes the
representation label $k$, establishing a direct connection between
collective dynamics and the underlying group structure of the many-body
system. This collapse highlights that although the microscopic excitation
processes are protocol-dependent, they all produce states within the
same SU$(1,1)$ manifold, characterized by a single effective parameter.
The agreement across different protocols and final trap frequencies
thus provides a direct and robust verification of the symmetry-protected
amplitude-energy-fluctuation relation.

\begin{figure}
\includegraphics[width=1\columnwidth]{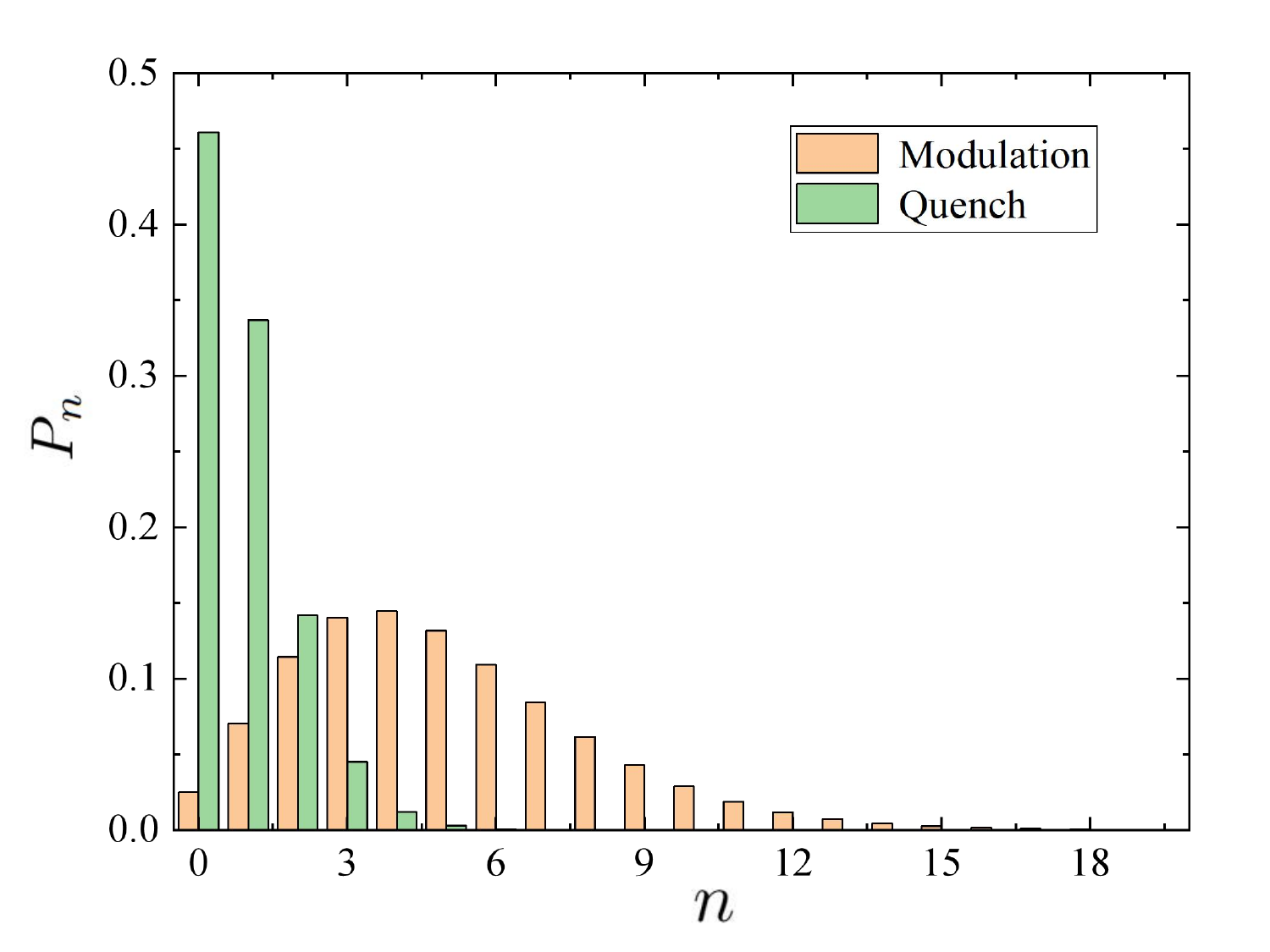}

\caption{(Color online) Transition probabilities $P_{n}$ to the quasiparticle
(tower) states $\left|n\right\rangle $ after excitation. Results
for resonant modulation and quench protocols are compared, showing
distinct microscopic excitation distributions. For resonant modulation,
the trap frequency is modulated for ten cycles with $\beta=0.1$;
for the quench protocol, the final frequency is $\omega=2\omega_{0}$.}

\label{fig:TransitionProbability}
\end{figure}

To gain microscopic insight into the excitation process, we examine
the transition probabilities from the initial quasiparticle vacuum
to the breathing-mode tower states, as shown in Fig.\ref{fig:TransitionProbability}.
These probabilities determine the full energy distribution after excitation
and take a universal form \citep{SM}
\begin{equation}
P_{n}=\frac{\left(n+2k-1\right)!}{n!\left(2k-1\right)!}\frac{\tanh^{2n}\left(\mathcal{S}_{\text{eff}}\right)}{\cosh^{4k}\left(\mathcal{S}_{\text{eff}}\right)},
\end{equation}
fully determined by a single parameter $\mathcal{S}_{\text{eff}}.$
This form follows from the SU$(1,1)$ structure of the dynamics and
applies to both quench and resonant modulation, with $\mathcal{S}_{\text{eff}}=v$
for a quench and $\mathcal{S}_{\text{eff}}=s$ for resonant driving.
Despite the distinct excitation mechanisms, all protocol dependence
is encode in $\mathcal{S}_{\text{eff}}$, reducing the many-body excitation
to a one-parameter description and leading directly to the universal
relation between energy fluctuation and breathing amplitude shown
in Fig.\ref{fig:UniversalRatio}.

\emph{\prlsection{Conclusions}}In conclusion, we have uncovered a
universal connection between energy fluctuations and collective dynamics
in scale-invariant quantum gases. The breathing-mode amplitude provides
a direct and quantitative probe of energy fluctuations, dictated by
the underlying SO$(2,1)$ dynamical symmetry. While different excitation
protocols populate the many-body spectrum in distinct ways, all resulting
states are governed by a single effective parameter, leading to an
exact and protocol-independent relation between microscopic fluctuations
and macroscopic observables. Remarkably, this relation follows solely
from symmetry and therefore holds universally for a broad class of
quantum systems governed by scale invariance and related dynamical
symmetry, ranging from conformal quantum mechanics and inverse-square
potential systems to critical quantum field theories and quantum gases
\citep{DeAlfaro1976C,Cardy1996S,Francesco1997C,Camblong2003A}. This
demonstrates that the apparent complexity of nonequilibrium many-body
dynamics can be strongly reduced by symmetry, enabling direct experimental
access to energy fluctuations without full spectral reconstruction.
Our work establishes a symmetry-based paradigm for probing quantum
fluctuations and opens a new route toward accessing nonequilibrium
thermodynamics in strongly interacting systems.
\begin{acknowledgments}
This work was supported by the National Key R\&D Program of China
(Grant No. 2022YFA1404102), the National Natural Science Foundation
of China (Grant Nos. U23A2073 and 12374250), and the Quantum Science
and Technology National Science and Technology Major Project (Grant
No. 2023ZD0300401).

Shi-Guo Peng and Jing Min contributed equally to the work.
\end{acknowledgments}

\clearpage
\onecolumngrid
\appendix

\section*{Supplementary Material:Breathing Modes as a Probe of Energy Fluctuations in a Unitary Fermi
Gas}
\makeatletter
\@removefromreset{equation}{section}
\makeatother

\setcounter{equation}{0}
\renewcommand{\theequation}{S\arabic{equation}}


\section{SO$\left(2,1\right)$ symmetry and Casimir invariant}

A spherically trapped unitary Fermi gas exhibits an emergent conformal
symmetry. The free-space Hamiltonian 
\begin{equation}
\hat{H}_{0}=\sum_{i}\frac{{\bf p}^{2}_{i}}{2m}+V_{\text{int}}\left(\left\{ {\bf r}_{i}\right\} \right),
\end{equation}
together with the dilatation operator 
\begin{equation}
\hat{D}=\frac{1}{2}\sum_{i}\left({\bf r}_{i}\cdot{\bf p}_{i}+{\bf p}_{i}\cdot{\bf r}_{i}\right),
\end{equation}
and the special conformal operator 
\begin{equation}
\hat{K}=\frac{1}{2}\sum_{i}mr^{2}_{i}
\end{equation}
form a closed Lie algebra
\begin{equation}
\left[\hat{K},\hat{H}_{0}\right]=i\hbar\hat{D},\quad\left[\hat{D},\hat{H}_{0}\right]=2i\hbar\hat{H}_{0},\quad\left[\hat{K},\hat{D}\right]=2i\hbar\hat{K},
\end{equation}
which realizes the SO$\left(2,1\right)$ dynamical symmetry of the
system. Here, ${\bf r}_{i}$ and ${\bf p}_{i}$ denote the coordinate
and momentum of the $i$th atom. In the presence of a harmonic trap
with frequency $\omega$, it is convenient to introduce the generators
\citep{Pitaevskii1997B}
\begin{equation}
\hat{L}_{0}=\frac{\hat{H}+\omega^{2}\hat{K}}{2\omega},\quad\hat{L}_{1}=\frac{\hat{H}-\omega^{2}\hat{K}}{2\omega},\quad\hat{L}_{2}=\frac{\hat{D}}{2},
\end{equation}
which satisfy
\begin{equation}
\left[\hat{L}_{0},\hat{L}_{1}\right]=i\hbar\hat{L}_{2},\quad\left[\hat{L}_{1},\hat{L}_{2}\right]=-i\hbar\hat{L}_{0},\quad\left[\hat{L}_{2},\hat{L}_{0}\right]=i\hbar\hat{L}_{1}.
\end{equation}
This is the standard SO$\left(2,1\right)$ algebra, isomorphic to
the Lie algebra of the 2D Lorentz group. Introducing ladder operators
\begin{equation}
\hat{L}_{\pm}=\frac{1}{\sqrt{2}}\left(\hat{L}_{1}\pm i\hat{L}_{2}\right),
\end{equation}
one obtains 
\begin{equation}
\left[\hat{L}_{0},\hat{L}_{\pm}\right]=\pm\hbar\hat{L}_{\pm},\quad\left[\hat{L}_{+},\hat{L}_{-}\right]=-\hbar\hat{L}_{0},
\end{equation}
which is the canonical SU$\left(1,1\right)$ form. 

The quadratic Casimir operator is given by \citep{Werner2006U},
\begin{equation}
\hat{C}=\hat{\mathcal{H}}^{2}-\left(2\omega\right)^{2}\left(\hat{L}_{+}\hat{L}_{-}+\hat{L}_{-}\hat{L}_{+}\right),\label{eq:54}
\end{equation}
which satisfies $\left[\hat{C},\hat{L}_{0,\pm}\right]=0$. It therefore
labels irreducible representation of the algebra and commutes with
the trapped Hamiltonian $\hat{\mathcal{H}}=\hat{H}_{0}+\omega^{2}\hat{K}$.
We consider simultaneous eigenstates,
\begin{equation}
\hat{\mathcal{H}}\left|\psi^{\left(\lambda\right)}_{\epsilon}\right\rangle =\epsilon\left|\psi^{\left(\lambda\right)}_{\epsilon}\right\rangle ,\quad\hat{C}\left|\psi^{\left(\lambda\right)}_{\epsilon}\right\rangle =\lambda\left|\psi^{\left(\lambda\right)}_{\epsilon}\right\rangle .
\end{equation}
Since $\left[\hat{C},\hat{L}_{\pm}\right]=0$, the ladder operators
act within a fixed-$\lambda$ sector. Their action shifts the energy
as
\begin{equation}
\hat{\mathcal{H}}\left(\hat{L}_{\pm}\left|\psi^{\left(\lambda\right)}_{\epsilon}\right\rangle \right)=\left(\epsilon\pm2\hbar\omega\right)\left(\hat{L}_{\pm}\left|\psi^{\left(\lambda\right)}_{\epsilon}\right\rangle \right),
\end{equation}
showing that $\hat{L}_{\pm}$ raise/lower the energy by exactly $2\hbar\omega$.
Assuming the lowest-energy state $\left|\psi^{\left(\lambda\right)}_{\epsilon_{g}}\right\rangle $
satisfying $\hat{L}_{-}\left|\psi^{\left(\lambda\right)}_{\epsilon_{g}}\right\rangle =0$,
the spectrum forms a ladder 
\begin{equation}
\left|\psi^{\left(\lambda\right)}_{\epsilon_{g}+n\left(2\hbar\omega\right)}\right\rangle \propto\hat{L}^{n}_{+}\left|\psi^{\left(\lambda\right)}_{\epsilon_{g}}\right\rangle .
\end{equation}
Defining the Bargmann index $k=\epsilon_{g}/2\hbar\omega$, which
characterizes the irreducible representation of SU$\left(1,1\right)$
and is fixed by the many-body state, one obtains
\begin{equation}
\lambda=\left(2\hbar\omega\right)^{2}k\left(k-1\right).
\end{equation}
The relevant many-body states can thus be labeled as $\left|k,n\right\rangle $
with 
\begin{equation}
\hat{L}_{0}\left|k,n\right\rangle =\hbar\left(n+k\right)\left|k,n\right\rangle ,\quad\hat{C}\left|k,n\right\rangle =\left(2\hbar\omega\right)^{2}k\left(k-1\right)\left|k,n\right\rangle .
\end{equation}
Assuming the standard discrete-series representation of SU$\left(1,1\right)$,
the ladder operators act as
\begin{eqnarray}
\hat{L}_{+}\left|k,n\right\rangle  & = & \frac{\hbar}{\sqrt{2}}\sqrt{\left(n+1\right)\left(n+2k\right)}\left|k,n+1\right\rangle ,\\
\hat{L}_{-}\left|k,n\right\rangle  & = & \frac{\hbar}{\sqrt{2}}\sqrt{n\left(n-1+2k\right)}\left|k,n-1\right\rangle ,
\end{eqnarray}
which follows from the SU$\left(1,1\right)$ algebra and the Casimir
constraint, and fully determines the breathing-mode tower.

\section{breathing quasiparticles}

Within a fixed $\lambda$ sector, the Bargamann index $k$ is fixed,
and the tower states can be labeled as $\left|n\right\rangle \equiv\left|k,n\right\rangle $.
Motivated by the ladder structure, we introduce bosonic operators
$\left\{ \hat{b}^{\dagger},\hat{b}\right\} $ satisfying $\left[\hat{b},\hat{b}^{\dagger}\right]=1,$with
\begin{equation}
\hat{b}^{\dagger}\left|n\right\rangle =\sqrt{n+1}\left|n+1\right\rangle ,\quad\hat{b}\left|n\right\rangle =\sqrt{n}\left|n-1\right\rangle ,
\end{equation}
and number operator $\hat{n}=\hat{b}^{\dagger}\hat{b}$. In this basis,
the SU$\left(1,1\right)$ generators take the form
\begin{equation}
\hat{L}_{0}=\hbar\left(\hat{n}+k\right),\quad\hat{L}_{+}=\hat{L}^{\dagger}_{-}=\frac{\hbar}{\sqrt{2}}\hat{b}^{\dagger}\sqrt{\hat{n}+2k}.\label{eq:2ndQuantization}
\end{equation}
The trapped Hamiltonian becomes
\begin{equation}
\hat{\mathcal{H}}=2\omega\hat{L}_{0}=\left(2\hbar\omega\right)\left(\hat{n}+k\right).
\end{equation}
This representation identifies the lowest-energy state $\left|0\right\rangle $
as a quasiparticle vacuum, with the tower states corresponding to
excitations of bosonic mode with energy spacing $2\hbar\omega$. The
breathing dynamics can thus be interpreted as the creation and annihilation
of collective quasiparticles associated with the SO$\left(2,1\right)$
symmetry.

\section{Time evolution in the quasiparticle representation}

The time-dependent Hamiltonian 
\[
\hat{\mathcal{H}}\left(t\right)=\hat{H}_{0}+\omega^{2}\left(t\right)\hat{K}
\]
 is a linear combination of SO$\left(2,1\right)$ generators. The
time-evolution operator therefore admits an exact disentangled form
\citep{Zhang2022Q,Ban1993D},
\begin{equation}
\hat{U}\left(t\right)=e^{\sqrt{2}\zeta_{+}\left(t\right)\hat{L}_{+}/\hbar}e^{\hat{L}_{0}\ln\left[1-\left|\zeta_{+}\left(t\right)\right|^{2}\right]/\hbar}e^{-\sqrt{2}\zeta^{*}_{+}\left(t\right)\hat{L}_{-}/\hbar}.
\end{equation}
The complex parameter $\zeta_{+}\left(t\right)$ satisfies the Riccati
equation
\begin{equation}
i\frac{d}{dt}\zeta_{+}\left(t\right)=\frac{1}{2}\alpha_{-}\left(t\right)+\alpha_{+}\left(t\right)\zeta_{+}\left(t\right)+\frac{1}{2}\alpha_{-}\left(t\right)\zeta^{2}_{+}\left(t\right),\label{eq:RiccatiEq}
\end{equation}
with 
\begin{equation}
\alpha_{\pm}\left(t\right)=\omega_{0}\left[1\pm\frac{\omega^{2}\left(t\right)}{\omega^{2}_{0}}\right].
\end{equation}
It is convenient to parameterize the evolution in terms of a displacement
operator
\begin{equation}
\hat{U}_{D}\left(\xi\right)=e^{\sqrt{2}\left(\xi\hat{L}_{+}-\xi^{*}\hat{L}_{-}\right)/\hbar},
\end{equation}
where
\begin{equation}
\xi\left(t\right)=s\left(t\right)e^{i\theta\left(t\right)},\quad\zeta_{+}\left(t\right)=e^{i\theta\left(t\right)}\tanh\left(s\right).
\end{equation}
In this form, the full many-body dynamics is equivalent to a displacement
(squeezing) transformation, and is characterized by a single complex
parameter $\xi\left(t\right)$.

We now express the dynamics in the quasiparticle representation defined
with respect to the initial trap frequency $\omega_{0}$. Since the
Casimir eigenvalue $\lambda$ is conserved, the evolution remains
within a fixed irreducible representation, and all operators can be
expressed in terms of the quasiparticle operators $\left\{ \hat{b}^{\dagger}_{0},\hat{b}_{0}\right\} $.
The time evolution of the annihilation operator is conveniently formulated
as
\begin{equation}
\hat{b}\left(t\right)=\hat{U}^{\dagger}_{D}\left(\xi\right)\hat{b}_{0}\hat{U}_{D}\left(\xi\right).
\end{equation}
The transformation follows from the SU$\left(1,1\right)$ rotation
of the generators,
\begin{equation}
\left[\begin{array}{c}
\hat{L}_{0}\left(t\right)\\
\hat{L}_{+}\left(t\right)\\
\hat{L}_{-}\left(t\right)
\end{array}\right]=\mathcal{M}_{D}\left(s,\theta\right)\left[\begin{array}{c}
\hat{L}_{0}\\
\hat{L}_{+}\\
\hat{L}_{-}
\end{array}\right],
\end{equation}
where
\begin{equation}
\mathcal{M}_{D}\left(s,\theta\right)=\left[\begin{array}{ccc}
\cosh\left(2s\right) & \frac{e^{i\theta}}{\sqrt{2}}\sinh\left(2s\right) & \frac{e^{-i\theta}}{\sqrt{2}}\sinh\left(2s\right)\\
\frac{e^{-i\theta}}{\sqrt{2}}\sinh\left(2s\right) & \cosh^{2}\left(s\right) & e^{-i2\theta}\sinh^{2}\left(s\right)\\
\frac{e^{i\theta}}{\sqrt{2}}\sinh\left(2s\right) & e^{i2\theta}\sinh^{2}\left(s\right) & \cosh^{2}\left(s\right)
\end{array}\right].
\end{equation}
Using the quasiparticle representation (\ref{eq:2ndQuantization})
together with their time-evolved counterparts, we obtain 
\begin{eqnarray}
\hat{n}\left(t\right)+k & = & \cosh\left(2s\right)\left(\hat{n}_{0}+k\right)+\frac{e^{i\theta}}{2}\sinh\left(2s\right)\hat{b}^{\dagger}_{0}\sqrt{\hat{n}_{0}+2k}+\frac{e^{-i\theta}}{2}\sinh\left(2s\right)\sqrt{\hat{n}_{0}+2k}\hat{b}_{0},\label{eq:TEvolutionOfQPO1}\\
\hat{b}^{\dagger}\left(t\right)\sqrt{\hat{n}\left(t\right)+2k} & = & e^{-i\theta}\sinh\left(2s\right)\left(\hat{n}_{0}+k\right)+\cosh^{2}\left(s\right)\hat{b}^{\dagger}_{0}\sqrt{\hat{n}_{0}+2k}+e^{-i2\theta}\sinh^{2}\left(s\right)\sqrt{\hat{n}_{0}+2k}\hat{b}_{0},\label{eq:TEvolutionOfQPO2}\\
\sqrt{\hat{n}\left(t\right)+2k}\hat{b}\left(t\right) & = & e^{i\theta}\sinh\left(2s\right)\left(\hat{n}_{0}+k\right)+e^{i2\theta}\sinh^{2}\left(s\right)\hat{b}^{\dagger}_{0}\sqrt{\hat{n}_{0}+2k}+\cosh^{2}\left(s\right)\sqrt{\hat{n}_{0}+2k}\hat{b}_{0}.\label{eq:TEvolutionOfQPO3}
\end{eqnarray}
These relations constitute a nonlinear displacement transformation
of the quasiparticle operators. In contrast to the linear Bogoliubov
transformations encountered in quadratic systems \citep{Xin2021R},
the present structure reflects the underlying SU$\left(1,1\right)$
algebra and the interacting nature of the many-body problem. The dynamics
therefore corresponds to a symmetry-constrained evolution within a
fixed representation, with all time dependence encoded in the parameters
$s\left(t\right)$ and $\theta\left(t\right)$.

\section{Quasiparticle number fluctuations}

We consider a unitary Fermi gas initially prepared in equilibrium
in a spherical harmonic trap with frequency $\omega_{0}$. The system
occupies the lowest-energy state of a fixed SO$\left(2,1\right)$
irreducible representation, corresponding to the quasiparticle vacuum
$\left|0;\omega_{0}\right\rangle $. Subsequent evolution driven by
a time-dependent harmonic confinement generates a displacement within
the same representation. 

The quasiparticle number is defined as
\begin{equation}
\mathcal{N}\left(t\right)=\left\langle \hat{n}\left(t\right)\right\rangle _{0}=\left\langle 0;\omega_{0}\right|\hat{n}\left(t\right)\left|0;\omega_{0}\right\rangle .
\end{equation}
Using the transformation derived above, 
\begin{equation}
\hat{n}\left(t\right)+k=\cosh\left(2s\right)\left(\hat{n}_{0}+k\right)+\frac{e^{i\theta}}{2}\sinh\left(2s\right)\hat{b}^{\dagger}_{0}\sqrt{\hat{n}_{0}+2k}+\frac{e^{-i\theta}}{2}\sinh\left(2s\right)\sqrt{\hat{n}_{0}+2k}\hat{b}_{0},
\end{equation}
together with $\hat{b}_{0}\left|0;\omega_{0}\right\rangle =0$, one
obtains
\begin{equation}
\mathcal{N}\left(t\right)=k\left[\cosh\left(2s\right)-1\right].
\end{equation}
The quasiparticle number fluctuation follows from 
\begin{equation}
\Delta\mathcal{N}\left(t\right)=\sqrt{\left\langle \hat{n}^{2}\left(t\right)\right\rangle _{0}-\left\langle \hat{n}\left(t\right)\right\rangle ^{2}_{0}}=\sqrt{\frac{k}{2}}\sinh\left(2s\right).
\end{equation}
It is convenient to express these results in terms of $\zeta_{+}\left(t\right)=e^{i\theta}\tanh\left(s\right)$,
for which
\begin{equation}
\cosh\left(2s\right)=\frac{1+\left|\zeta_{+}\right|^{2}}{1-\left|\zeta_{+}\right|^{2}},\;\sinh\left(2s\right)=\frac{2\left|\zeta_{+}\right|}{1-\left|\zeta_{+}\right|^{2}}.
\end{equation}
The parameter $\zeta_{+}\left(t\right)$ satisfies the Riccati equation
(\ref{eq:RiccatiEq}). Once $\zeta_{+}\left(t\right)$ is determined,
both $\mathcal{N}$ and $\Delta\mathcal{N}\left(t\right)$ follow
directly. The fluctuations scale proportionally with the quasiparticle
number at large excitation, reflecting the collective nature of the
SO$\left(2,1\right)$ dynamics.

\section{Amplitude-energy-fluctuation relation}

We now derive the relation between the energy fluctuation and the
breathing-mode amplitude within the quasiparticle framework. After
an excitation of duration $t_{1}$, the system evolves in a static
harmonic trap with frequency $\omega$. The post-excitation state
is generated from the initial ground state as
\begin{equation}
\left|\psi_{1}\right\rangle =\hat{U}_{D}\left(\xi\right)\left|0;\omega_{0}\right\rangle ,
\end{equation}
where $\hat{U}_{D}\left(\xi\right)$ is the SU$\left(1,1\right)$
displacement operator during the excitation. To describe the subsequent
free evolution, it is convenient to express all observables in the
quasiparticle basis associated with the final trap frequency $\omega$.
In this basis, the Hamiltonian takes the diagonal form 
\begin{equation}
\hat{\mathcal{H}}\left(\omega\right)=\left(2\hbar\omega\right)\left(\hat{n}+k\right),
\end{equation}
so that the total energy and its fluctuation are entirely determined
by the quasiparticle number and its fluctuation
\begin{equation}
E=\left(2\hbar\omega\right)\left(\mathcal{N}+k\right),\quad\Delta E=\left(2\hbar\omega\right)\Delta\mathcal{N}.
\end{equation}
The state $\left|\psi_{1}\right\rangle $, however, is defined with
respect to the initial trap frequency $\omega_{0}$. The relation
between the two quasiparticle bases is governed by a scale transformation
$\hat{U}_{S}\left(v\right)=e^{iv\hat{D}/\hbar}$ with $e^{2v}=\omega/\omega_{0}$.
Consequently, the Hamiltonian at different trap frequencies are related
by $\hat{\mathcal{H}}\left(\omega\right)=\hat{U}_{S}\left(v\right)\hat{\mathcal{H}}\left(\omega_{0}\right)\hat{U}^{\dagger}_{S}\left(v\right)$.
This transformation induces a hyperbolic rotation in the SU$\left(1,1\right)$
generators. Expressing the generators in terms of quasiparticle operators,
one finds that the scale transformation mixes creation and annihilation
operators in a nonlinear manner. In particular, we have
\begin{equation}
\left[\begin{array}{c}
\hat{L}_{0}\left(\omega\right)\\
\hat{L}_{+}\left(\omega\right)\\
\hat{L}_{-}\left(\omega\right)
\end{array}\right]=\mathcal{M}_{S}\left(v\right)\left[\begin{array}{c}
\hat{L}_{0}\left(\omega_{0}\right)\\
\hat{L}_{+}\left(\omega_{0}\right)\\
\hat{L}_{-}\left(\omega_{0}\right)
\end{array}\right],
\end{equation}
where
\begin{equation}
\mathcal{M}_{S}\left(v\right)=\left[\begin{array}{ccc}
\cosh\left(2v\right) & \frac{1}{\sqrt{2}}\sinh\left(2v\right) & \frac{1}{\sqrt{2}}\sinh\left(2v\right)\\
-\frac{1}{\sqrt{2}}\sinh\left(2v\right) & \cosh^{2}\left(v\right) & \sinh^{2}\left(v\right)\\
-\frac{1}{\sqrt{2}}\sinh\left(2v\right) & \sinh^{2}\left(v\right) & \cosh^{2}\left(v\right)
\end{array}\right].
\end{equation}
The quasiparticle operators are transformed as
\begin{eqnarray}
\hat{n}+k & = & \cosh\left(2v\right)\left(\hat{n}_{0}+k\right)-\frac{\sinh\left(2v\right)}{2}\hat{b}^{\dagger}_{0}\sqrt{\hat{n}_{0}+2k}-\frac{\sinh\left(2v\right)}{2}\sqrt{\hat{n}_{0}+2k}\hat{b}_{0},\label{eq:ScaleTranOFQPO1}\\
\hat{b}^{\dagger}\sqrt{\hat{n}+2k} & = & -\sinh\left(2v\right)\left(\hat{n}_{0}+k\right)+\cosh^{2}\left(v\right)\hat{b}^{\dagger}_{0}\sqrt{\hat{n}_{0}+2k}+\sinh^{2}\left(v\right)\sqrt{\hat{n}_{0}+2k}\hat{b}_{0},\label{eq:ScaleTranOFQPO2}\\
\sqrt{\hat{n}+2k}\hat{b} & = & -\sinh\left(2v\right)\left(\hat{n}_{0}+k\right)+\sinh^{2}\left(v\right)\hat{b}^{\dagger}_{0}\sqrt{\hat{n}_{0}+2k}+\cosh^{2}\left(v\right)\sqrt{\hat{n}_{0}+2k}\hat{b}_{0}.\label{eq:ScaleTranOFQPO3}
\end{eqnarray}
This shows that a change of the trap frequency is equivalent to a
nonlinear squeezing transformation of quasiparticles dictated by the
SO$\left(2,1\right)$ symmetry.

Combining this scale transformation with displacement evolution, the
quasiparticle number in the final basis can be evaluated as
\begin{equation}
\mathcal{N}+k=k\left[\cosh\left(2s\right)\cosh\left(2v\right)-\sinh\left(2s\right)\sinh\left(2v\right)\cos\theta\right],
\end{equation}
while the corresponding fluctuation is
\begin{equation}
\Delta\mathcal{\mathcal{N}}=\sqrt{\frac{k}{2}}\left\{ \left[\sinh\left(2s\right)\cosh\left(2v\right)-\cosh\left(2s\right)\sinh\left(2v\right)\cos\theta\right]^{2}+\sinh^{2}\left(2v\right)\sin^{2}\theta\right\} ^{1/2}.
\end{equation}
To connect with a directly observable quantity, we consider the expectation
value of the special conformal operator $\hat{K}$, which determines
the cloud size. During the free evolution, it exhibits harmonic oscillation
at frequency $2\omega$, 
\begin{equation}
\left\langle \hat{K}\right\rangle \left(t\right)=\frac{\hbar}{\omega}\left(\mathcal{N}+k\right)-\frac{\hbar k}{\omega}\mathcal{A}\cos\left(2\omega\tau-\delta\right)
\end{equation}
with amplitude
\begin{equation}
\mathcal{A}=\left\{ \left[\sinh\left(2s\right)\cosh\left(2v\right)\cos\theta-\cosh\left(2s\right)\sinh\left(2v\right)\right]^{2}+\left[\sinh\left(2s\right)\sin\theta\right]^{2}\right\} ^{1/2}
\end{equation}
and phase shift
\begin{equation}
\tan\delta=\frac{\sinh\left(2s\right)\sin\theta}{\sinh\left(2s\right)\cosh\left(2v\right)\cos\theta-\cosh\left(2s\right)\sinh\left(2v\right)}.
\end{equation}
A direct comparison between the above expressions shows that the amplitude
and the quasiparticle fluctuation satisfy the exact identity
\begin{equation}
\mathcal{A}=\sqrt{\frac{2}{k}}\Delta\mathcal{N}.
\end{equation}
Combining this with $\Delta E=\left(2\hbar\omega\right)\Delta\mathcal{N}$,
we finally obtain
\begin{equation}
\frac{\Delta E/\hbar\omega}{\mathcal{A}/a^{2}_{\text{ho}}}=\frac{1}{\sqrt{2k}}.
\end{equation}
This result shows that the mean cloud size is determined by the average
quasiparticle number, while the oscillation amplitude is governed
solely by its fluctuation. As a consequence, the breathing-mode amplitude
provides a direct and quantitative probe of energy fluctuations, with
their ratio entirely by the Bargmann index $k$, independent of excitation
protocols and microscopic details. This relation establishes a direct
fluctuation--response correspondence for the breathing mode: the
oscillation amplitude encodes the quantum energy uncertainty in a
universal and dimensionless manner.

\section{Transition probabilities for breathing-mode excitation}

We derive the transition probabilities to the breathing-mode tower
states for both sudden quenches and resonant modulations, and show
that they take a universal form fixed by the underlying SO$\left(2,1\right)$
structure. We first consider a sudden quench of the trap frequency
from $\omega_{0}$ to $\omega$, with the system initially prepared
in the ground state $\left|0;\omega_{0}\right\rangle $. Expanding
the state in the eigenbasis of the post-quench Hamiltonian, one has
\begin{equation}
\left|0;\omega_{0}\right\rangle =\sum_{n}c_{n}\left|n;\omega\right\rangle ,
\end{equation}
where 
\begin{equation}
\hat{\mathcal{H}}\left(\omega\right)\left|n;\omega\right\rangle =2\hbar\omega\left(n+k\right)\left|n;\omega\right\rangle .
\end{equation}
The coefficients $c_{n}$ are determined by the lowest-energy condition
satisfied by the initial state
\begin{equation}
\sqrt{\hat{n}_{0}+2k}\hat{b}_{0}\left|0;\omega_{0}\right\rangle =0.
\end{equation}
Using the scale transformation that relates quasiparticle operators
at frequencies $\omega_{0}$ and $\omega$, i.e., Eqs.(\ref{eq:ScaleTranOFQPO1})-(\ref{eq:ScaleTranOFQPO3}),
this condition can be expressed in the $\omega$ basis, leading to
a recursion relation
\begin{equation}
\sinh\left(2v\right)\left(n+k\right)c_{n}+\sinh^{2}\left(v\right)c_{n-1}\sqrt{\left(n-1+2k\right)n}+\cosh^{2}\left(v\right)c_{n+1}\sqrt{\left(n+2k\right)\left(n+1\right)}=0.
\end{equation}
Solving this recursion yields the transition probability
\begin{equation}
P_{n}=\left|c_{n}\right|^{2}=\frac{\left(n+2k-1\right)!}{n!\left(2k-1\right)!}\frac{\tanh^{2n}\left(v\right)}{\cosh^{4k}\left(v\right)}.
\end{equation}

We now consider excitation by resonant modulation of the trap frequency.
The time evolution is governed by an SU$\left(1,1\right)$ displacement
operator, such that the state at the end of the integer-cycle modulation
can be written as $\left|\psi_{1}\right\rangle =\hat{U}_{D}\left(\xi\right)\left|0;\omega_{0}\right\rangle $
with $\xi=se^{i\theta}$. Using the disentangled form of the SU$\left(1,1\right)$
transformation together with the ladder structure of the generators,
one obtains
\begin{equation}
\left|\psi_{1}\right\rangle =\sum_{n}\sqrt{\frac{\left(n+2k-1\right)!}{n!\left(2k-1\right)!}}\frac{\tanh^{n}\left(s\right)}{\cosh^{2k}\left(s\right)}e^{in\theta}\left|n;\omega_{0}\right\rangle .
\end{equation}
In this basis, the expression coefficients directly determine the
transition probabilities,
\begin{equation}
P_{n}=\frac{\left(n+2k-1\right)!}{n!\left(2k-1\right)!}\frac{\tanh^{2n}\left(s\right)}{\cosh^{4k}\left(s\right)}.
\end{equation}
Restricting to the physically relevant situation of integer-cycle
modulation, both excitation protocols lead to the identical probability
distribution
\begin{equation}
P_{n}=\frac{\left(n+2k-1\right)!}{n!\left(2k-1\right)!}\frac{\tanh^{2n}\left(\mathcal{S}_{\text{eff}}\right)}{\cosh^{4k}\left(\mathcal{S}_{\text{eff}}\right)},
\end{equation}
where $\mathcal{S}_{\text{eff}}=v$ for a quench and $\mathcal{S}_{\text{eff}}=s$
for resonant modulation. In this case, the evolution reduces to a
pure SU$\left(1,1\right)$ displacement within a fixed irreducible
representation, and the transition probabilities are governed by a
single effective parameter. 
%
%
%
%
%

\end{document}